\documentclass[fleqn]{article}
\usepackage{spconf,amsmath,graphicx}

\usepackage{cite}
\usepackage{url}
\usepackage{bm}
\usepackage{hyperref}
\usepackage{xspace}
\usepackage{verbatim}
\usepackage{multirow}
\usepackage{booktabs}
\usepackage[table,xcdraw]{xcolor}
\usepackage{array}
\usepackage{amsmath,amssymb}
\usepackage{tabularx}
\newcolumntype{Y}{>{\centering\arraybackslash}X}

\usepackage{multirow}
\usepackage{makecell}
\usepackage{multicol}
\usepackage{enumitem, paralist}

\newcommand\tabref[1]{Table~\ref{#1}}
\newcommand\secref[1]{Section~\ref{#1}}

\def\x{{\bm x}}
\def\y{{\bm y}}
\def\z{{\bm z}}
\def\L{{\cal L}}

\title{Nonparallel High-Quality Audio Super Resolution \\with Domain Adaptation and Resampling CycleGANs}
\name{Reo Yoneyama$^{1*}$\thanks{*Work performed during an internship at LINE Corporation.}, Ryuichi Yamamoto$^{2}$, and Kentaro Tachibana$^{2}$}
\address{$^{1}$Nagoya University, Japan, $^{2}$LINE Corp., Japan.}

\begin{document}
\fontsize{9.4}{11.5}\selectfont
\setlength{\mathindent}{4pt}
\maketitle
\begin{abstract}
Neural audio super-resolution models are typically trained on low- and high-resolution audio signal pairs. Although these methods achieve highly accurate super-resolution if the acoustic characteristics of the input data are similar to those of the training data, challenges remain: the models suffer from quality degradation for out-of-domain data, and paired data are required for training. To address these problems, we propose Dual-CycleGAN, a high-quality audio super-resolution method that can utilize unpaired data based on two connected cycle consistent generative adversarial networks (CycleGAN). Our method decomposes the super-resolution method into domain adaptation and resampling processes to handle acoustic mismatch in the unpaired low- and high-resolution signals. The two processes are then jointly optimized within the CycleGAN framework. Experimental results verify that the proposed method significantly outperforms conventional methods when paired data are not available. Code and audio samples are available from \url{https://chomeyama.github.io/DualCycleGAN-Demo/}.
\end{abstract}

\section{Introduction}
\label{sec:intro}
Audio super-resolution (SR) (also called bandwidth extension) is a technique used to predict a high-resolution (HR) audio signal (e.g., at 48~kHz) from a low-resolution (LR) signal (e.g., at 16~kHz). 
The potential applications of audio SR are wide-ranging and include refining old audio or video recordings, enhancement of speech generated by a text-to-speech (TTS) systems \cite{hifigan+}, and restoring frequency-bands missing from low-resolution data in recognition tasks; other uses include automatic speech recognition \cite{asr-app1, asr-app2, asr-app3}, speaker recognition \cite{sr-app1}, speaker identification, and speaker verification \cite{sv-app1, sv-app2}. 


Recent neural audio SR models based on generative adversarial networks (GANs)\cite{mugan, hifigan+}, normalizing flows\cite{wsrglow}, and diffusion probabilistic models\cite{nuwave, nuwave2} were all trained on the pairs of corresponding LR and HR audio signals.
Although these methods achieve accurate SR when the acoustic characteristics of the input data are similar to those of training data, they suffer from quality degradation for the out-of-domain data.
This is problematic in the real-world applications to which SR is applied outside the training domain, e.g., speech recorded in different acoustic environments and synthesized speech from TTS systems.

One possible way to solve this problem is to train SR models using data from the target domain to familiarize the model with it.
However, conventional SR methods\cite{mugan, nugan, nuwave, nuwave2, wsrglow, hifigan+, nvsr, tunet} require pairs of LR and HR data (i.e., parallel data), and HR audio signals from the target domain are generally unavailable. 
Therefore, it is inherently difficult to solve these out-of-domain issues using conventional methods, and thus its applications are limited.

To address these problems, we propose Dual-CycleGAN, a high-quality nonparallel audio SR method based on two connected cycle consistent generative adversarial networks (CycleGAN) \cite{CycleGAN}.
In contrast to the conventional methods, Dual-CycleGAN uses both nonparallel and parallel data and learns an implicit mapping between the LR and HR signals, even when HR signals are unavailable for the target domain.
Furthermore, we decompose SR into domain adaptation and resampling (i.e., up and downsampling processes) to solve the mismatch between the acoustic characteristics of the unpaired LR and HR signals. 
These two processes are modeled using two connected CycleGANs and are optimized via a two-stage training approach for stable optimization: joint pre-training and fine-tuning.
Experimental results show that the Dual-CycleGAN outperforms the conventional methods under two challenging SR conditions: 1) SR across different recording environments and 2) SR on TTS-generated speech.

Note that there is a concurrent study that aims to solve for domain adaptation and SR jointly based on two time-domain CycleGANs for speaker verification\cite{sv-app2}. 
The researchers investigated several optimization strategies and applied their method successfully to upsampling of telephone signals from 8~kHz to 16~kHz.
We tackle more challenging SR conditions, i.e., upsampling of 16~kHz signals to 48~kHz for high-quality SR, which is essential for applications such as enhancement of TTS-generated speech.

\section{Proposed method}
\label{sec:proposed-method}

\subsection{Overview}
\begin{figure}[tb]
\begin{center}
    \vspace{-1mm}
    \includegraphics[width=0.97\columnwidth]{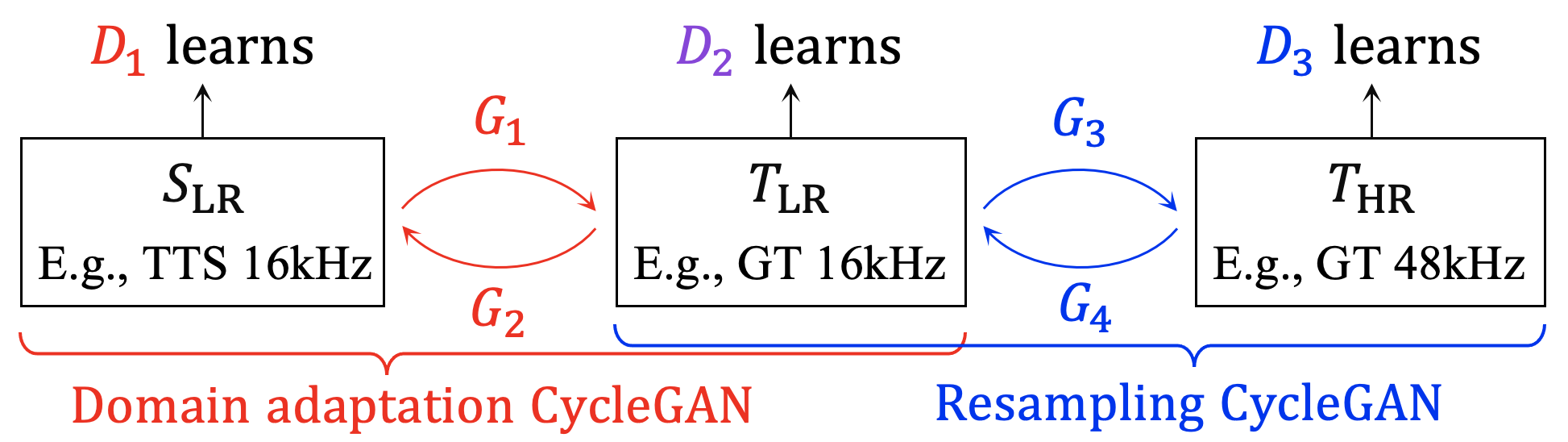}
    \vspace{-3mm}
    \caption{
    \fontsize{9.4}{11.5}\selectfont Overview of the proposed Dual-CycleGAN.
    }
    \label{fig:dual-cyclegan-1}
    \end{center}
\vspace{-6mm}
\end{figure}

Figure~\ref{fig:dual-cyclegan-1} shows an overview of the proposed Dual-CycleGAN.
Our method comprises two components: domain adaptation and resampling, where each is based on a CycleGAN\cite{CycleGAN}.
We denote the generator and discriminator pairs for the domain adaptation CycleGAN by $\{G_1, G_2\}$ and $\{D_1, D_2\}$, and the corresponding pairs for the resampling CycleGAN are $\{G_3, G_4\}$ and $\{D_2, D_3\}$, respectively. 
Note that the two CycleGANs share $D_2$ to save their parameters.

The resampling CycleGAN learns the conversion between LR and HR speech waveforms in the same domain $T$, where parallel data are available. 
We define the LR and HR domains of $T$ as $T_{\text{LR}}$ and $T_{\text{HR}}$, respectively.
To generate a high quality HR speech waveform from LR speech from the different domain $S_{\text{LR}}$, the domain adaptation CycleGAN learns mappings between $S_{\text{LR}}$ and $T_{\text{LR}}$.
The inference is performed as the composition mapping $S_{\text{LR}} \rightarrow T_{\text{LR}} \rightarrow T_{\text{HR}}$ by $G_3(G_1(\cdot))$.

\subsection{Two-stage optimization}
\subsubsection{Joint pre-training}
In the joint pre-training process, the two conversions $S_{\text{LR}} \leftrightarrow T_{\text{LR}}$ and $T_{\text{LR}} \leftrightarrow T_{\text{HR}}$ are trained simultaneously following the general CycleGAN training approach based on least-squares GANs~\cite{lsgan}. 
In the domain adaptation CycleGAN, $D_1$ learns to identify the generated samples from $G_2$ as $\textit{fake}$ and the ground-truth (GT) samples $\x \in
S_{\text{LR}}$ as $\textit{real}$, while $D_2$ learns to identify the generated samples from $G_1$ as $\textit{fake}$ and the GT samples $\z \in T_{\text{LR}}$ as $\textit{real}$. In the resampling CycleGAN, $D_2$ learns to identify the generated samples from $G_4$ as $\textit{fake}$ and the GT samples $\z \in T_{\text{LR}}$ as $\textit{real}$, while $D_3$ learns to identify the generated samples from $G_3$ as $\textit{fake}$, and the GT samples $\y \in T_{\text{HR}}$ as $\textit{real}$. 

The generators learn to minimize the weighted sum of three losses: the adversarial loss $\L_{\text{adv}}$, the cycle-consistency loss $\L_{\text{cyc}}$, and the identity mapping loss $\L_{\text{idt}}$. These losses are formulated as follows:
\begin{flalign}
    \begin{split}
    \L_{\text{adv}}&(G_1, G_2, G_3, G_4 ; D_1, D_2, D_3) 
    \\ =& ~ \mathbb{E}_{\x, \z} \left[ (1 - D_1(G_2(\z)))^2 + (1 - D_2(G_1(\x)))^2 \right]
    \\ +& ~ \mathbb{E}_{\y, \z} \left[ (1 - D_2(G_4(\y)))^2 + (1 - D_3(G_3(\z)))^2 \right], \label{eq:pre_adv}
    \end{split}
    \vspace{-2mm}
\end{flalign}
\begin{flalign}
    \begin{split}
    &\L_{\text{cyc}}(G_1, G_2, G_3, G_4)
    \\ &= \mathbb{E}_{\x, \z} \left[ ~ \lVert \x - G_2(G_1(\x)) \rVert_1 + \lVert \z - G_1(G_2(\z)) \rVert_1 ~ \right]
    \\ &+\mathbb{E}_{\y, \z} \left[ ~ \lVert \y - G_3(G_4(\y)) \rVert_1 + \lVert \z - G_4(G_3(\z)) \rVert_1 ~ \right],  \label{eq:pre_cyc}
    \end{split}
    \vspace{-2mm}
\end{flalign}
\begin{flalign}
    \begin{split}
    \L_{\text{idt}}(& G_1, G_2, G_3, G_4) 
    \\ &= \mathbb{E}_{\x, \z} \left[ ~ \lVert \x - G_2(\x) \rVert_1 + \lVert \z - G_1(\z) \rVert_1 ~ \right]
    \\ &+ \mathbb{E}_{\y, \z} \left[ ~ \lVert \y - G_3(\z) \rVert_1 + \lVert \z - G_4(\y) \rVert_1 ~ \right].  \label{eq:pre_idt}
    \end{split}
    \vspace{-1mm}
\end{flalign}
We set the weights of $\L_{\text{adv}}$, $\L_{\text{cyc}}$ and $\L_{\text{idt}}$ empirically to 1, 10, and 10, respectively.

The discriminators learn to identify the fake and real samples by learning to minimize the sum of two adversarial losses: the fake loss $\L_{\text{fake}}$ and the real loss $\L_{\text{real}}$.
\begin{flalign}
    \vspace{-1mm}
    \begin{split}
    \L_{\text{fake}}(& D_1, D_2, D_3 ; G_1, G_2, G_3, G_4) 
    \\ &= \mathbb{E}_{\x, \z} \left[ D_1(G_2(\z))^2 + D_2(G_1(\x))^2 / ~ 2 ~ \right]
    \\ &+ \mathbb{E}_{\y, \z} \left[ D_2(G_4(\y))^2 / ~ 2 + D_3(G_3(\z))^2 ~ \right], \label{eq:pre_fake}
    \end{split}
    \vspace{-2mm}
\end{flalign}
\begin{flalign}
    &\L_{\text{real}}(D_1, D_2, D_3) \label{eq:pre_real}
    \\ &= \mathbb{E}_{\x, \y, \z} \left[ (1 - D_1(\x))^2 + (1 - D_2(\z))^2 + (1 - D_3(\y))^2 \right]. \nonumber
    \vspace{-1mm}
\end{flalign}
As described in \secref{ssec:fine}, we can optimize the entire network in an end-to-end manner, but joint pre-training makes the entire training process stable.

\vspace{-1mm}
\subsubsection{Fine-tuning}
\label{ssec:fine}
The fine-tuning process enables a direct optimization of the composite mappings $S_{\text{LR}} \rightarrow T_{\text{LR}} \rightarrow T_{\text{HR}}$ and its inverse $T_{\text{HR}} \rightarrow T_{\text{LR}} \rightarrow S_{\text{LR}}$ to ease the effect of the domain shift between the $\textit{real}$ and $\textit{fake}$ samples of $T_{\text{LR}}$, induced by the imperfect domain adaptation performed by $G_1$.

We rewrite three of the loss functions used in the joint pre-training process. 
In Equations~\eqref{eq:pre_adv} and \eqref{eq:pre_fake}, the generated samples $G_2(\z)$ and $G_3(\z)$ are replaced by $G_2(G_4(\y))$ and $G_3(G_1(\x))$, respectively.
Furthermore, Equation \eqref{eq:pre_cyc} is replaced with a new cycle-consistency loss that considers the entire cycle between the four generators and is defined as follows:
\begin{flalign}
    \vspace{-1mm}
    \begin{split}
    \L_{\text{cyc}}(& G_1, G_2, G_3, G_4)
    \\ &= ~ \mathbb{E}_{\x} \left[ ~ \lVert \x - G_2(G_4(G_3(G_1(\x)))) \rVert_1 ~ \right]
    \\ &+ ~ \mathbb{E}_{\x} \left[ ~ \lVert G_1(\x) - G_4(G_3(G_1(\x))) \rVert_1 ~ \right]
    \\ &+ ~ \mathbb{E}_{\y} \left[ ~ \lVert \y - G_3(G_1(G_2(G_4(\y)))) \rVert_1 ~ \right]
    \\ &+ ~ \mathbb{E}_{\y} \left[ ~ \lVert G_4(\y) - G_1(G_2(G_4(\y))) \rVert_1 ~ \right]. \label{eq:joint_cyc}
    \end{split}
    \vspace{-1mm}
\end{flalign}
The fine-tuning is performed as the same manner as the joint pre-training process, except that the three newly defined loss functions are used in this case.

\subsection{Perceptual cycle consistency}
\label{perceptual cycle-consistency}
It is possible to use the waveform-domain L1 distance to compute the cycle-consistency loss functions\cite{wavecyclegan, wavecyclegan2}. 
However, because there are multiple possible HR waveforms for what corresponds to the same LR waveform due to the ill-posed nature of SR, the waveform-domain L1 loss would make it difficult for the CycleGANs to learn cycle-consistent functions that should be bijections\cite{CycleGAN}.
To address this issue, we compute a perceptually meaningful cycle-consistency loss based on the mel-spectrogram.
Specifically, we replace the waveform-domain L1 loss in Equations~\eqref{eq:pre_cyc} and \eqref{eq:joint_cyc} with the mel-scale time-frequency domain L1 loss. 
Similarly, we replace the identity mapping loss in Equation~\eqref{eq:pre_idt} with the proposed perceptual loss.
Because the mel-spectral matching brings flexibility to the high-frequency bands, the CycleGANs can learn conversions that consider the ill-posed nature of SR implicitly.

\subsection{Network architecture}
All generators of the Dual-CycleGAN have the same architecture, which is based on the WaveCycleGAN2 generator\cite{wavecyclegan2}.
We replaced the first and last linear projection layers of the base architecture with a one-dimensional convolutional neural network (CNN) followed by gated linear units (GLUs) \cite{glu}.
To perform resampling, we add a sinc-based interpolation with 151 kernel size before the first layer only for $G_3$ and $G_4$.

Several studies have shown the importance of combining multiple types of discriminators for speech generation tasks \cite{wavecyclegan2, gan-vocoder, univnet}. Inspired by these studies, we use multi-domain discriminators for the waveform and the log amplitude linear spectral domains.
For the waveform domain discriminator, we use the Parallel WaveGAN's one~\cite{pwg}.
For the spectral domain discriminator, we use the multi-band grouped discriminators of the NU-GAN \cite{nugan}. 
Each sub-discriminator of the NU-GAN captures band-wise spectral features independently according to its unique grouping size, thus helping the generators learn all the frequency bands effectively.
To balance the loss ratio of the two domains of the discriminators, we use a fully connected layer to summarize the multiple outputs from the spectral discriminators into one scalar value.

\section{Experimental Evaluations}
\label{sec:experimental evaluations}
To evaluate the performance of the proposed Dual-CycleGAN, we conducted two experiments: SR across different recording environments and SR on TTS-generated speech. 
In this section, we first describe the details of the dataset and model configurations. 
We then describe the results from the two experiments.

\subsection{Experimental setups}

\subsubsection{Database}
We used the two publicly available databases LJ~Speech \cite{ljspeech17} and VCTK\cite{vctk} (mic2, version 0.92), as the nonparallel and parallel datasets, respectively.
LJ~Speech contains approximately 24 hours of data recorded by a single English female speaker. 
These audio signals were downsampled from 22.05~kHz to 16~kHz, and were then used as data for the domain $S_\text{LR}$.
VCTK consists of approximately 44 hours of clean speech signals recorded by 110 English speakers. 
The VCTK audio signals were provided at 48~kHz.
We performed high-pass filtering with a cutoff frequency of 70 Hz to remove the low-frequency noise.
Each audio signal was normalized to -26 dB.
To prepare paired LR and HR signals for VCTK, audio signals at 16~kHz were constructed by performing downsampling using the librosa \cite{librosa} resample function.
Then, the paired data of the 16~kHz and 48~kHz signals were used as samples of $S_\text{LR}$ and $T_\text{LR}$.
We split the datasets into training, validation, and test sets. 
For LJ~Speech, we used 100 and 250 utterances for the validation and test sets, respectively, and the rest were used for training. 
For VCTK, we split the datasets for each speaker: we used 200 and 400 utterances for the validation and test sets, respectively, and the rest were kept for training.

\subsubsection{Model details}
\label{ssec:model}
The proposed Dual-CycleGAN model was trained for the first 400~K iterations with the joint pre-training approach and then fine-tuned for 200~K iterations, using the Adam optimizer \cite{adam} ($\epsilon=10^{-8}$, $\beta=[0.5, 0.999]$). 
The identity mapping loss was used for up to 100~K iterations. 
We set the minibatch size to four, the length of each audio clip to 12~K time samples (0.75~s at 16~kHz), and the initial learning rates were set to 0.0002 and 0.0001 for the generators and discriminators, respectively. 
These learning rates were reduced by half after every 200~K iterations.
We applied weight normalization \cite{salimans2016weight} to all convolutional layers and clipped the gradient norms at ten for all networks. 

As the baselines, the HiFi-GAN+ \cite{hifigan+} and WSRGlow \cite{wsrglow} models were used. 
Because these baseline models cannot use unpaired datasets, we only used the VCTK dataset for training.
HiFi-GAN+ consists of a feed-forward WaveNet \cite{wavenet} architecture trained using a GAN-based deep feature loss.
We trained the HiFi-GAN+ model using an open-source implementation\footnote{\url{https://github.com/brentspell/hifi-gan-bwe/}}.
The model was trained for 1000~K steps, followed by an additional 100~K steps of joint training using the additive noise data augmentation \cite{hifigan+} with the DNS challenge dataset \cite{reddy20_interspeech}.
WSRGlow uses a Glow-based generative model to perform audio SR \cite{kingma2018glow}.
Because the official implementation \footnote{\url{https://github.com/zkx06111/WSRGlow}} does not support SR from 16~kHz to 48~kHz, we re-implemented the WSRGlow model based on the official version.
The hyperparameters were the same as those described in the original paper \cite{wsrglow}, with the exception of those related to the resampling factor.

We also prepared a single CycleGAN~\cite{CycleGAN} based nonparallel SR model to investigate the effectiveness of the explicit domain adaptation. 
The model architecture was the same as that of the Dual-CycleGAN, except that its generators have three more layers than the Dual-CycleGAN to match the number of parameters approximately to that of the Dual-CycleGAN.
To calculate the identity mapping loss, which requires parallel data, we trained the model on the VCTK parallel data for the first 100~K iterations.
We then trained it for 500~K iterations using all data.

\subsection{SR across different recording environments}
\label{ssec:exp1}

\subsubsection{Comparison with baselines}
\label{ssec:exp1_comp}
We conducted subjective preference tests\footnote{
Although previous works~\cite{hifigan+,wsrglow} used objective tests based on spectral distances, we do not adopt them for lack of target speech.
} to evaluate the performance of the proposed method.
We asked 20 subjects to evaluate their preference from a randomly selected pair of two samples from the test set.
Specifically, 48~kHz samples were generated from LJ~Speech's 16~kHz speech using the trained SR models, and they were then used to create pairs of samples.
Note that each pair consisted of samples of the proposed and baseline methods.
We also included the recorded samples at 22.05~kHz for comparison.
The subjects were asked to listen to the entire utterances.
In total, 200 samples for each method pair were evaluated.

\begin{table}[t]
\vspace{-2mm}
\caption{
\fontsize{9.4}{11.5}\selectfont Preference scores for SR across different recording environments.}
\vspace{1mm}
\label{table:results}
\scalebox{1}{
\begin{tabular}{lccc}
\Xhline{2\arrayrulewidth}
\small Baseline & \small Baseline wins & \small Neutral & \small Proposed wins \\
\Xhline{2\arrayrulewidth}
\small HiFi-GAN+ & \small $31.2$ \% & \small $26.7$ \% & \small $\textbf{42.1}$ \% \\
\small WSRGlow & \small $21.2$ \% & \small $15.8$ \% & \small $\textbf{63.0}$ \% \\
\small CycleGAN & \small $22.6$ \% & \small $32.1$ \% & \small $\textbf{45.3}$ \% \\
\Xhline{2\arrayrulewidth}
\small Recordings & \small $\textbf{47.9}$ \% & \small $33.0$ \% & \small $19.1$ \% \\
\Xhline{2\arrayrulewidth}
\end{tabular}
}
\vspace{-4mm}
\end{table}

\begin{table}[t]
\vspace{-2mm}
\caption{
\fontsize{9.4}{11.5}\selectfont Preference scores for the ablation study.}
\vspace{1mm}
\label{table:ablation}
\scalebox{1}{
\begin{tabular}{lccc}
\Xhline{2\arrayrulewidth}
\small Baseline & \small Baseline wins & \small Neutral & \small Proposed wins \\
\Xhline{2\arrayrulewidth}
\small w/o fine tune & \small $19.5$ \% & \small $52.0$ \% & \small $\textbf{28.5}$ \% \\
\small w/o mel-loss & \small $2.0$ \% & \small $5.0$ \% & \small $\textbf{93.0}$ \% \\
\Xhline{2\arrayrulewidth}
\end{tabular}
}
\vspace{-3mm}
\end{table}

\tabref{table:results} shows the results of the preference tests, which can be summarized as follows:
\begin{inparaenum}[(1)]
\item the proposed method outperformed HiFi-GAN+ and WSRGlow significantly.
Although HiFi-GAN+ and WSRGlow tended to generate audible artifacts because of the acoustic mismatch between the training and the inference time (i.e., the models were trained on VCTK but tested on LJ~Speech), our method was able to generate clearer speech by using both nonparallel and parallel data during the training process.
\item The importance of explicit domain adaptation was verified by comparing the single CycleGAN-based method with the proposed method.
\end{inparaenum}
Note that we observed that the proposed method is inferior to the 22.05~kHz recorded speech. 
We hypothesize that these results occurred because some of the VCTK data contain audible noise.
We believe that the quality gap between the recorded and generated speech can be reduced if higher quality clean data from the $T_{\text{HR}}$ domain are used.

\subsubsection{Ablation study}
We performed an ablation study to investigate the effectiveness of the proposed fine-tuning process and the mel-spectral-domain cyclic and identity mapping loss functions.
We prepared two models for comparison; one model was jointly pre-trained for 600~K iterations, and the other model was trained with both the cycle-consistency and identity mapping losses in the waveform domain.
Preference tests were performed as per \secref{ssec:exp1_comp} under the same test conditions.

\tabref{table:ablation} shows the preference test results.
These results show that the fine-tuning procedure improved the speech quality from the jointly pre-trained model.
Furthermore, a comparison between the proposed model and the proposed model without the mel-spectral losses confirmed the importance of the mel-spectral cycle consistency and identity mapping losses.

\subsection{SR on TTS-generated speech}

\begin{table}[t]
\vspace{-2mm}
\begin{center}         
\caption{
\fontsize{9.4}{11.5}\selectfont Sound quality mean opinion score (MOS) test result with 95\% confidence interval for SR on TTS-generated speech.
}
\vspace{1mm}
\label{tab:tts_gen}
\scalebox{1}{
{
\begin{tabular}{lcc}
\Xhline{2\arrayrulewidth}
\small System & \small MOS $\uparrow$ \\
\hline
\small VITS (16~kHz) & \small $3.97 \pm 0.11$ \\
\small VITS (48~kHz) & \small $4.54 \pm 0.08$ \\
\small HiFi-GAN+ (16~kHz $\rightarrow$ 48~kHz) & \small $4.23 \pm 0.10$ \\
\small WSRGlow (16~kHz $\rightarrow$ 48~kHz) & \small $4.23 \pm 0.11$ \\
\small \textbf{Ours} (16~kHz $\rightarrow$ 48~kHz) & \small $\textbf{4.51} \pm \textbf{0.09}$ \\
\Xhline{2\arrayrulewidth}
\small Recordings (48~kHz) & \small $4.65 \pm 0.07$  \\
\Xhline{2\arrayrulewidth}
\end{tabular}}    
}
\end{center}
\vspace{-3mm}
\end{table}

\subsubsection{TTS setup}
To evaluate the effectiveness of the proposed method further, we performed additional experiments for SR on TTS-generated speech.
In these experiments, we used the VCTK dataset only. 
We defined the 16~kHz TTS-generated speech as samples of the domain $S_\text{LR}$. $T_\text{LR}$ and $T_\text{HR}$ were the domains of the 16~kHz and 48~kHz recorded samples of VCTK, respectively. 
The test set comprises two seen male and two seen female speakers (designated p351, p361, p363, and p364) and the four unseen speakers (designated p362, p374, p376, and s5), where the numbers of utterances by each speaker are identical. 

For the TTS system, we used a multi-speaker version of a high-quality end-to-end TTS model (i.e., VITS~\cite{vits}).
We trained the model on the 16~kHz VCTK dataset for 1000~K steps using the AdamW optimizer\cite{loshchilov2018decoupled}.
We also prepared another VITS model that was trained on the same corpus with the 48~kHz sampling rate. 
The detailed model architecture and training setups were the same as those in the original paper\cite{vits}, although the upsampling factors were adjusted for the 16~kHz and 48~kHz sampling rates.
To enhance the TTS-generated speech, the proposed Dual-CycleGAN was trained using both recorded and synthetic data, while the HiFi-GAN+ and WSRGlow models were trained using the recorded data only.
The details of the model architecture and training configurations were the same as those in \secref{ssec:model}.

\subsubsection{Subjective listening tests}
To analyze the performances of TTS systems with the proposed SR method, we performed five scaled mean opinion score (MOS) tests.
We evaluated three TTS systems that were enhanced by HiFi-GAN+, WSRGlow, and our proposed model. 
We also evaluated two VITS-based TTS systems and 48~kHz clean recordings for comparison.
We asked 20 subjects to judge the sound quality of 20 randomly selected speech samples from the test set.
The subjects listened to the entire utterances of each sample.

\tabref{tab:tts_gen} shows the results of these MOS tests.
These results showed that
\begin{inparaenum}[(1)]
    \item our proposed TTS system obtained the best score of 4.51 among all the enhanced TTS systems, significantly outperforming the two baseline SR methods; and
    \item a TTS system trained on LR speech (i.e., VITS trained on 16~kHz speech) can be improved significantly to match the quality of a system trained on HR data (i.e., VITS trained on 48~kHz speech).
\end{inparaenum}

\section{Conclusion}
This paper introduced the Dual-CycleGAN, a nonparallel audio SR method to perform high-quality SR across different domains. 
We demonstrated that the Dual-CycleGAN outperformed the baseline SR methods in two challenging situations in which parallel data were unavailable.
The intended future direction is to develop an any-to-one domain adaptation method to handle input speech from arbitrary domains.
Moreover, investigation of cross-lingual SR will be valuable for low-resource languages without sufficient amounts of studio-quality data.

\newpage

\bibliographystyle{IEEEbib}
{
\fontsize{9.15}{11.35}\selectfont
\bibliography{main}

\begin{thebibliography}{10}

\bibitem{hifigan+}
J.~Su, Y.~Wang, A.~Finkelstein, et~al.,
\newblock ``{Bandwidth Extension is All You Need},''
\newblock in {\em Proc. ICASSP}, 2021, pp. 696--700.

\bibitem{asr-app1}
M.~L. Seltzer and A.~Acero,
\newblock ``{Training Wideband Acoustic Models Using Mixed-Bandwidth Training
  Data for Speech Recognition},''
\newblock {\em IEEE Trans. on Audio, Speech, and Lang. Process.}, vol. 15, no.
  1, pp. 235--245, 2007.

\bibitem{asr-app2}
D.~Haws and X.~Cui,
\newblock ``{Cyclegan Bandwidth Extension Acoustic Modeling for Automatic
  Speech Recognition},''
\newblock in {\em Proc. ICASSP}, 2019, pp. 6780--6784.

\bibitem{asr-app3}
X.~Li, V.~Chebiyyam, and K.~Kirchhoff,
\newblock ``{Speech Audio Super-Resolution for Speech Recognition},''
\newblock in {\em Proc. Interspeech}, 2019, pp. 3416--3420.

\bibitem{sr-app1}
H.~Yamamoto, K.~A. Lee, K.~Okabe, et~al.,
\newblock ``{Speaker Augmentation and Bandwidth Extension for Deep Speaker
  Embedding},''
\newblock in {\em Proc. Interspeech}, 2019, pp. 406--410.

\bibitem{sv-app1}
H.~Miyamoto, S.~Shiota, and H.~Kiya,
\newblock ``{Application of Bandwidth Extension with No Learning to Data
  Augmentation for Speaker Verification},''
\newblock in {\em Odyssey}, 2020, pp. 446--450.

\bibitem{sv-app2}
S.~Kataria, J.~Villalba, L.~Moro-Velázquez, et~al.,
\newblock ``{Joint domain adaptation and speech bandwidth extension using
  time-domain GANs for speaker verification},''
\newblock in {\em Proc. Interspeech}, 2022, pp. 615--619.

\bibitem{mugan}
S.~Kim and V.~Sathe,
\newblock ``Bandwidth extension on raw audio via generative adversarial
  networks,''
\newblock {\em arXiv preprint arXiv:1903.09027}, 2019.

\bibitem{wsrglow}
K.~Zhang, Y.~Ren, C.~Xu, et~al.,
\newblock ``{WSRGlow: A Glow-Based Waveform Generative Model for Audio
  Super-Resolution},''
\newblock in {\em Proc. Interspeech}, 2021, pp. 1649--1653.

\bibitem{nuwave}
J.~Lee and S.~Han,
\newblock ``{NU-Wave: A Diffusion Probabilistic Model for Neural Audio
  Upsampling},''
\newblock in {\em Proc. Interspeech}, 2021, pp. 1634--1638.

\bibitem{nuwave2}
S.~Han and J.~Lee,
\newblock ``{NU-Wave 2: A General Neural Audio Upsampling Model for Various
  Sampling Rates},''
\newblock in {\em Proc. Interspeech}, 2022, pp. 4401--4405.

\bibitem{nugan}
R.~Kumar, K.~Kumar, V.~Anand, Y.~Bengio, and A.~Courville,
\newblock ``{NU-GAN: High resolution neural upsampling with GAN},''
\newblock {\em arXiv preprint arXiv:2010.11362}, 2020.

\bibitem{nvsr}
H.~Liu, W.~Choi, X.~Liu, et~al.,
\newblock ``{Neural Vocoder is All You Need for Speech Super-resolution},''
\newblock in {\em Proc. Interspeech}, 2022, pp. 4227--4231.

\bibitem{tunet}
N.~Viet~Anh, A.~Nguyen, and A.~Khong,
\newblock ``{TUNet: A Block-Online Bandwidth Extension Model Based On
  Transformers And Self-Supervised Pretraining},''
\newblock in {\em Proc. ICASSP}, 2022, pp. 161--165.

\bibitem{CycleGAN}
J.-Y. Zhu, T.~Park, P.~Isola, et~al.,
\newblock ``{Unpaired Image-to-Image Translation using Cycle-Consistent
  Adversarial Networks},''
\newblock in {\em Proc. ICCV}, 2017, pp. 2223--2232.

\bibitem{lsgan}
X.~Mao, Q.~Li, H.~Xie, et~al.,
\newblock ``{Least Squares Generative Adversarial Networks},''
\newblock in {\em Proc. ICCV}, 2017, pp. 2794--2802.

\bibitem{wavecyclegan}
K.~Tanaka, T.~Kaneko, N.~Hojo, and H.~Kameoka,
\newblock ``Synthetic-to-natural speech waveform conversion using
  cycle-consistent adversarial networks,''
\newblock in {\em Proc. SLT}, 2018, pp. 632--639.

\bibitem{wavecyclegan2}
K.~Tanaka, H.~Kameoka, T.~Kaneko, and N.~Hojo,
\newblock ``{WaveCycleGAN2: Time-domain Neural Post-filter for Speech Waveform
  Generation},''
\newblock {\em arXiv preprint arXiv:1904.02892}, 2019.

\bibitem{glu}
Y.~N. Dauphin, A.~Fan, M.~Auli, et~al.,
\newblock ``Language modeling with gated convolutional networks,''
\newblock in {\em Proc. ICML}, 2017, pp. 933--941.

\bibitem{gan-vocoder}
J.~You, D.~Kim, G.~Nam, G.~Hwang, and G.~Chae,
\newblock ``{GAN Vocoder: Multi-Resolution Discriminator Is All You Need},''
\newblock in {\em Proc. Interspeech}, 2021, pp. 2177--2181.

\bibitem{univnet}
W.~Jang, D.~Lim, J.~Yoon, et~al.,
\newblock ``{UnivNet: A Neural Vocoder with Multi-Resolution Spectrogram
  Discriminators for High-Fidelity Waveform Generation},''
\newblock in {\em Proc. Interspeech}, 2021, pp. 2207--2211.

\bibitem{pwg}
R.~Yamamoto, E.~Song, and J.-M. Kim,
\newblock ``{Parallel WaveGAN}: A fast waveform generation model based on
  generative adversarial networks with multi-resolution spectrogram,''
\newblock in {\em Proc. ICASSP}, 2020, pp. 6199--6203.

\bibitem{ljspeech17}
K.~Ito and L.~Johnson,
\newblock ``{The LJ Speech Dataset},''
  \url{https://keithito.com/LJ-Speech-Dataset/}, 2017.

\bibitem{vctk}
J.~Yamagishi, C.~Veaux, K.~MacDonald, et~al.,
\newblock ``{CSTR VCTK Corpus: English Multi-speaker Corpus for CSTR Voice
  Cloning Toolkit (version 0.92)},'' 2019.

\bibitem{librosa}
B.~McFee, A.~Metsai, M.~McVicar, et~al.,
\newblock ``librosa/librosa: 0.9.1,'' Feb. 2022.

\bibitem{adam}
D.~Kingma and J.~Ba,
\newblock ``{Adam: A Method for Stochastic Optimization},''
\newblock in {\em Proc. ICLR}, 2015.

\bibitem{salimans2016weight}
T.~Salimans and D.~P. Kingma,
\newblock ``{Weight normalization: A simple reparameterization to accelerate
  training of deep neural networks},''
\newblock in {\em Proc. NIPS}, 2016, pp. 901--909.

\bibitem{wavenet}
A.~v.~d. Oord, S.~Dieleman, H.~Zen, et~al.,
\newblock ``{WaveNet: A Generative Model for Raw Audio},''
\newblock in {\em Proc. SSW}, 2016.

\bibitem{reddy20_interspeech}
C.~K. Reddy, V.~Gopal, R.~Cutler, et~al.,
\newblock ``{The INTERSPEECH 2020 Deep Noise Suppression Challenge: Datasets,
  Subjective Testing Framework, and Challenge Results},''
\newblock in {\em Proc. Interspeech}, 2020, pp. 2492--2496.

\bibitem{kingma2018glow}
D.~P. Kingma and P.~Dhariwal,
\newblock ``{Glow: Generative Flow with Invertible 1x1 Convolutions},''
\newblock in {\em Proc. NeurIPS}, 2018, vol.~31.

\bibitem{vits}
J.~Kim, J.~Kong, and J.~Son,
\newblock ``{Conditional Variational Autoencoder with Adversarial Learning for
  End-to-End Text-to-Speech},''
\newblock in {\em Proc. ICML}, 2021, pp. 5530--5540.

\bibitem{loshchilov2018decoupled}
I.~Loshchilov and F.~Hutter,
\newblock ``{Decoupled Weight Decay Regularization},''
\newblock in {\em Proc. ICLR}, 2019.

\end{thebibliography}
}

\end{document}